\documentclass[epj,referee]{svjour}
\usepackage{graphics}
\usepackage{amsmath}
\begin{document}
\title{All-optical switching in metamaterial with high structural symmetry}
\subtitle{Bistable response of nonlinear double-ring planar
metamaterial}
\author{Vladimir R. Tuz\inst{1,2} \and Sergey L.
Prosvirnin\inst{1,
2}
}                     
\offprints{}          
\institute{Institute of Radioastronomy, National Academy of Sciences
of Ukraine, 4, Krasnoznamennaya st., Kharkiv 61002, Ukraine \and
School of Radio Physics, Karazin Kharkiv National University, 4,
Svobody Square, Kharkiv 61077, Ukraine}
\date{Received: date / Revised version: date}
%
\abstract{ We argue the possibility of realization of a
polarization-insensitive all-optical switching in a planar
metamaterial composed of a 4-fold periodic array of two concentric
metal rings placed on a substrate of nonlinear material. It is
demonstrated that a switching may be achieved between essentially
different values of transmission near the resonant frequency of the
high-quality-factor Fano-shape trapped-mode excitation.
} 
\maketitle
\section{Introduction}
\label{intro}

Optical bistability is an area of modern optics which is evolving
rapidly. The great attention to this phenomenon is due to the
possibility of its practical application to design optical switches,
limiters, transistors and diodes. Typically, in bistable devices,
the nonlinear medium is placed inside an optical cavity, just as is
done in lasers, but unlike the latter they are passive bistable
devices whose operating conditions are easy to control because the
light propagation is controlled with light \cite{gibbs}. A classical
example of the bistable device is a Fabry-Perot interferometer,
filled with a Kerr nonlinear material. In this case, the resonator
provides feedback, which is essential to obtain a multivalued
intensity at the structure's output. However, in such a system, both
relatively strong light power and/or large enough volume of
nonlinear optical material are generally needed to achieve a
sizeable nonlinear response.

To overcome these drawbacks, photonic crystal microcavities
\cite{soljacic} and quantum well structures \cite{chen} were
proposed to enhance the nonlinear effects as well as to reduce the
material volume. With the assistance of surface plasmon polaritons
to the effects of confining and enhancing the local optical field
intensity, optical bistability has also been shown numerically in
different metal nanostructures such as surface plasmon polaritonic
crystals \cite{wurtz}, subwavelength gratings that consisted of
infinitely long slits in metal slab \cite{min}, \cite{kochetova},
etc. In all these cases, the excitation of high-quality-factor
resonances in the systems is provided to obtain efficient switching.

A promising way to produce optical switching in compact devices can
be found also in using planar metamaterials. It is known that the
planar metamaterials can create an environment equivalent to a
resonant cavity. As usual such structures are composed of metallic
elements in the form of symmetrical split-ring resonators
\cite{sbrien}, \cite{sobrien}. They are resonant because of an
internal capacitance and inductance within each element. Most often
such structures are investigated in order to obtain a negative
refraction index over a finite frequency range, although their
nonlinear optical response was also studied
\cite{pendry}-\cite{brien}. Unfortunately a considerable
disadvantage of such structures in the context of obtaining optical
switching consists in their low quality factor.

Nevertheless, exceptionally strong and narrow resonances are
possible in planar metamaterials via engaging trapped modes
\cite{zouhdi}. The quality factor of the system depends on ratio of
power of stored energy to power of radiation and dissipation losses.
If we suppose infinitesimal dissipation losses in the structure, the
trapped modes correspond to real eigenvalues (i.e., real resonant
frequencies) of the relevant boundary value problem due to a special
geometry of metal elements. In the regime of quasi-trapped mode
excitation of an actual structure, the field is strongly localized
to the structure plane and resonant transmission and reflection have
a large quality factor due to very small radiation of
electromagnetic energy in a comparison with stored one.

Typically, metamaterials which can bear trapped modes consist of
identical subwavelength metallic inclusions structured in the form
of asymmetrically split rings \cite{zouhdi}, \cite{fedotov}, split
squares \cite{koch}, \cite{khardikov} or their complementary pattern
\cite{samson}, \cite{khardikov}. These elements are arranged
periodically and placed on a thin dielectric substrate. In the
structure of such kind, the high-quality-factor current oscillations
with the lowest total emission losses appear when all currents in
the metallic elements of a periodic cell oscillate in antiphase. The
characteristic feature of such metamaterials is the dependence of
their spectra on the polarization and the angle of incidence of
input waves. In \cite{prosvirnin}, \cite{kawakatsu}, the
polarization insensitive structure configurations also were
proposed. One such structure consists of a planar array with a
periodic cell element that consists of two concentric rings
(double-ring (DR) structure).

In the DR-structure, the trapped-modes excitation is exclusively
controlled by the difference in circumferences of the inner and
outer rings and does not depend on the polarization state of the
normally incident electromagnetic radiation. In contrast to the
asymmetrically split rings metamaterials, DR-structure is a system,
where mutual coupling between elements of neighboring cells is weak
and the response of the entire array is practically a direct sum of
the DR individual contributions. As a consequence, the
electromagnetic response of the DR-metamaterial is weakly dependent
on the angle of incidence \cite{papasimakis}. Remarkably, at the
trapped-mode resonance the electromagnetic energy is confined to a
very small region between the rings, where the energy density
reaches substantially high values. This makes the response of the
metamaterial operating in the trapped-mode regime extremely
sensitive to the dielectric properties of the substrate. This
feature can be used for enhancing optical nonlinear response in the
nanoscaled version of the metamaterial \cite{tuz}.

The goal of this paper is to show promising use of planar
metamaterials that bear the trapped-mode resonances to obtain
all-optical switching. Especially, we argue the possibility of
realization of a polarization-insensitive all-optical switching in
planar metamaterial designed on the basis of double-ring array
placed on a substrate of nonlinear dielectric in the regime of
trapped-mode excitation.

\section{Problem statement and solution}
\label{sec:1}

\begin{figure}
\resizebox{0.9\columnwidth}{!}{%
  \includegraphics{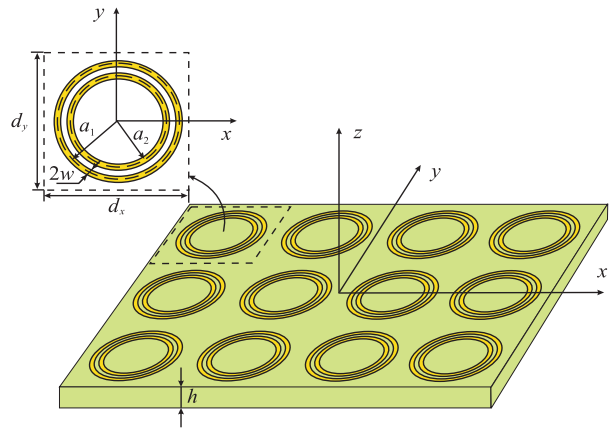}
} \caption{(Color online) Fragment of the planar metamaterial and
its elementary unit cell.} \label{fig:fig1}
\end{figure}

The square unit cell ($d=d_x=d_y$) of the structure under study
consists of one DR (Fig.~\ref{fig:fig1}). The radii of the outer and
inner rings are fixed at $a_1/d=0.36$ and $a_2/d=0.29$,
respectively. The width of both the metal rings is $2w/d=0.05$. The
array is placed on a nonlinear dielectric substrate with thickness
$h/d=0.2$. As a suitable variant of substrate material we propose
semiconductors appropriate in the mid-IR region of wavelength, for
example, InSb or InSb with some impurities such as As, Tl, Bi, P
\cite{razeghi}. From the experimental results
\cite{miller}-\cite{smith}, we expect that the Kerr-nonlinear part
of the refractive index of these materials in the mid-IR range is
between $10^{-4}$~cm$^2$/kW and $10^{-2}$~cm$^2$/kW.

As the excitation field a normally incident plane monochromatic wave
of a frequency $\omega$ and an amplitude $A$ is selected. We suppose
that the intensity of the incident field is enough for the
nonlinearity to become apparent, i.e., it is about 1~kW/cm$^2$.

In the set of our previous works \cite{zouhdi}, \cite{fedotov},
\cite{prosvirnin}-\cite{tuz} the method of moments
\cite{sprosvirnin} was used to calculate the response of different
planar metamaterials in the microwave range when the amplitude of
the incident wave is small (linear case). This numerical method
involves solving an integral equation for the surface currents
induced in the metallic pattern by the incident electromagnetic
wave, then calculating the scattered fields produced by the currents
as a superposition of partial spatial waves. In the framework of
this method it is implied that the metallic pattern is a thin
perfect conductor.

On the other hand in \cite{khardikov}, on the basis of the
pseudo-spectral time-domain algorithm, the optical response of the
metamaterial in the IR range was calculated taking into account the
strong dissipation and dispersion of the metal permittivity of the
elements. It was revealed that the trapped-mode resonances are well
observed and have high Q-factor in this range, and the obtained
results are in good agreements with the ones evaluated with the
method of moments down to the mid-IR region.

Thus in the present paper we will use the method of moments to
calculate the magnitude of the current $J$ along the single ring of
the DR element, the reflection $r$ and transmission $t$
coefficients. They can be determined in the form:
\begin{equation}\label{eq:linear}
J=J(\omega,\varepsilon),\quad t=t(\omega,\varepsilon),\quad
r=r(\omega, \varepsilon).
\end{equation}

Next we suppose that the structure substrate is a Kerr nonlinear
dielectric whose permittivity $\varepsilon$ depends on the intensity
of electric field $I_{in}$ inside it. Under rigorous consideration,
the nonlinear permittivity $\varepsilon$ of the substrate is
inhomogeneous. The permittivity riches its maximum value directly
under the metallic pattern and along the rings this permittivity is
also different. Nevertheless, as mentioned above, at the
trapped-mode resonance, the electromagnetic energy is confined to a
very small region between the rings and the crucial influence of the
permittivity on the system properties occurs in this place.
Therefore, the approximation based on the transmission line theory
can be used here to estimate the field intensity between the rings.
According to this theory, conductive rings are considered as two
wires with a distance $b$ between them. Along these wires the
currents flow in opposite directions. Thus the electric field
strength is defined as $$E_{in}=V/b,$$ where $V=Z J$ is the line
voltage, $b=a_1-a_2-2w$, $J$ is the magnitude of current which flows
along the DR-element, and $Z$ is the impedance of line. The
impedance is determined at the resonant dimensionless frequency
$\ae_0=d/\lambda_0$,
$$Z=60\frac{l\ae_0}{dC_0},$$ where $l=\pi(a_1+a_2)/2$, and
$$C_0=\frac{1}{4}\ln\left[\frac{p}{2w}+\sqrt{\left(\frac{p}{2w}\right)^2-1}\right]$$
is the capacity in free space per unit length of line, $p=a_1-a_2$.
From this model it follows that the electric field strength between
the rings is directly proportional to the current magnitude $J$.
Since the unit cell is small in comparison with the wavelength, the
current magnitude $J$ can be substituted with its value averaged
along the metallic ring, $\Bar J$.

Thus, from this model it follows that the electric field strength
between the rings is directly proportional to the average current
magnitude $\Bar J$, and the nonlinear equation on the average
current magnitude in the metallic pattern is obtained in the form
\begin{equation}\label{eq:nonlinear}
\Bar J=A \cdot \Bar J(\omega, \varepsilon_1+\varepsilon_2I_{in}(\Bar
J)).
\end{equation}

The incident field magnitude $A$ is a parameter of equation
(\ref{eq:nonlinear}). At a fixed frequency $\omega$, the solution of
this equation is the average current value which is dependent on the
magnitude of the incident field, $\Bar J=\Bar J(A)$.

On the basis of the current $\Bar J(A)$ found by a numerical
solution of equation (\ref{eq:nonlinear}), the value of the
permittivity of the nonlinear substrate
$\varepsilon=\varepsilon_1+\varepsilon_2{I_{in}(A)}$ is determined
and the reflection and transmission coefficients (\ref{eq:linear})
are calculated
\begin{equation}\label{eq:coeff}
t=t(\omega, \varepsilon_1+\varepsilon_2I_{in}(A)),\quad r=r(\omega,
\varepsilon_1+\varepsilon_2I_{in}(A)),
\end{equation}
as the functions of the frequency and magnitude of the incident
field.

It should be noted here that our treatment of nonlinearity in the
planar metamaterial differs from the ones considered earlier in
\cite{pendry}-\cite{brien} where some effective medium parameters
were introduced to obtain the nonlinear double negative materials.

\section{Numerical results}
\label{sec:3}

\begin{figure}
\resizebox{0.9\columnwidth}{!}{%
  \includegraphics{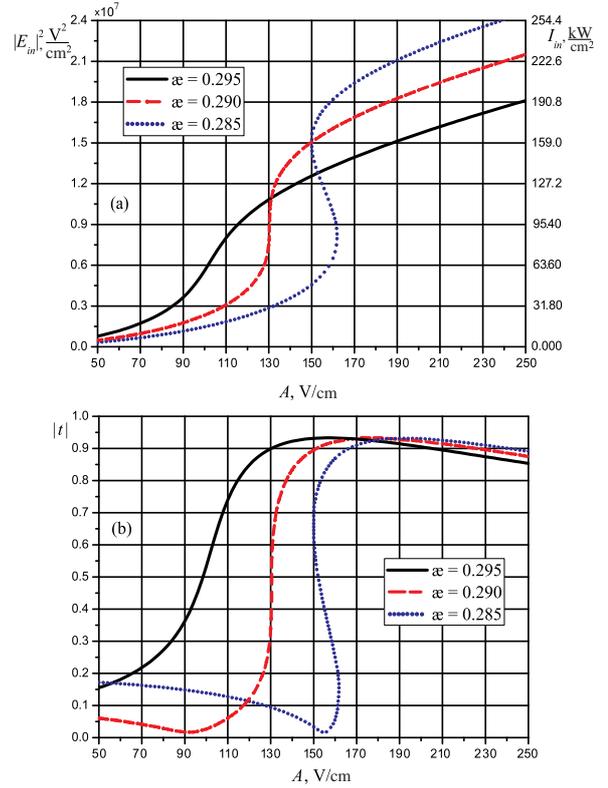}
} \caption{(Color online) (a) The inner intensity and (b) the
magnitude of the transmission coefficient versus the incident field
magnitude in the case of the nonlinear permittivity of the
substrate. For this and further calculations we use the substrate
parameters $\varepsilon_1=4.1+0.02i$ and $\varepsilon_2=5\times
10^{-3}$ cm$^2$/kW. The value of dimensionless frequency
$\ae=d/\lambda$ is chosen a bit lower to the frequency of the
trapped-mode resonance ($\ae_0=0.3$).} \label{fig:fig2}
\end{figure}

In the case of the nonlinear permittivity of substrate, dependences
of the inner intensity and the transmission coefficient magnitude
versus the incident field magnitude $I_{in}=I_{in}(A)$ are typical
and have the form of hysteresis (Fig.~\ref{fig:fig2}). Such form of
curves of the transmission coefficient magnitude is studied quite
well \cite{gibbs} and can be understood from the following
considerations. Suppose that the trapped mode resonant frequency is
slightly higher than the incident field frequency. As the intensity
of the incident field rises, the magnitude of currents on the metal
elements increases. This leads to increasing the field strength
inside the substrate and its permittivity as well. As a result, the
frequency of the resonant mode decreases and shifts toward the
frequency of incident wave, which, in turn, enhances further the
coupling between the current modes and the inner field intensity in
the nonlinear substrate. This positive feedback increases the slope
of the rising edge of the transmission spectrum, as compared to the
linear case. As the frequency extends beyond the resonant mode
frequency, the inner field magnitude in the substrate decreases and
the permittivity goes back toward its linear level, and this
negative feedback keeps the resonant frequency close to the incident
field frequency. As a result, at a certain intensity of the incident
field, the transmission coefficient stepwise changes its value from
small to large level.

\begin{figure}
\resizebox{0.9\columnwidth}{!}{%
  \includegraphics{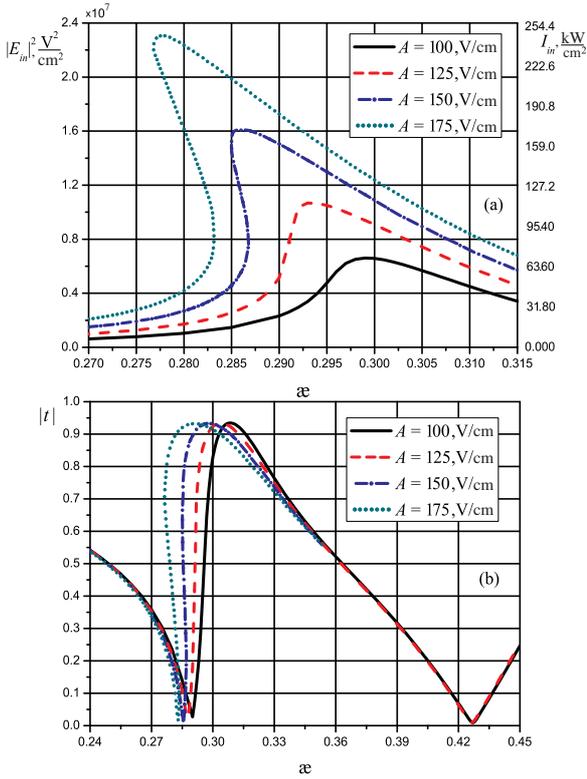}
} \caption{(Color online)  Frequency dependences of (a) the inner
intensity and (b) the magnitude of the transmission coefficient in
the case of the nonlinear permittivity ($\varepsilon_1=4.1+0.02i$
and $\varepsilon_2=5\times 10^{-3}$ cm$^2$/kW) of the substrate.}
\label{fig:fig3}
\end{figure}

The frequency dependences of the transmission coefficient magnitude
also manifests discontinuous switching from small to large level
with frequency increasing (Fig.~\ref{fig:fig3}). This switching
appears closely to the resonant frequency of the trapped-mode
excitation. The main peculiarity of the observed resonance is that
this trapped-mode resonance in DR-structure has the Fano-shape
rather than the Lorentzian one, as is the characteristic of
Fabry-Perot cavities. The Fano-type resonance appears as a result of
the interference between a high-quality resonance and a much
smoother, continuum-like spectrum and typically exhibits a sharp
asymmetric line shape with the transmission coefficients varying
from 0 to 1 over a very narrow frequency range
\cite{miroshnichenko}. Such form of resonance is very suitable to
obtain great amplitude of switching since there are gently sloping
bands of the high reflection and transmission before and after the
resonant frequency.

\section{Conclusion}
\label{sec:4}

In conclusion, a planar DR nonlinear metamaterial, which bears a
high-quality-factor Fano-shape trapped-mode resonance, is promising
object for a realization of a polarization-insensitive all-optical
switching.

This work was supported by the National Academy of Sciences of
Ukraine under the Program "Nanotechnologies and Nanomaterials",
Project No.~1.1.3.17.


\end{document}